\documentclass[10pt]{article}

\usepackage{times}
\usepackage{graphicx}
\usepackage{amsmath,amssymb}
\usepackage{multirow}
\usepackage{caption}
\usepackage{threeparttable}

\newenvironment{sciabstract}{%
\begin{quote} }
{\end{quote}}

\newcounter{lastnote}

\title{Concept of an Experimental Ultra Long-wavelength Radio Array}

\author
{Linjie~Chen,$^{1\ast}$ Yihua~Yan,$^{1,2}$\\
\\
\footnotesize{$^{1}$Key Laboratory of Solar Activity, National Astronomical Observatories, Chinese Academy of Sciences, Beijing 100012, China}\\
\footnotesize{$^{2}$School of Astronomy and Space Sciences, University of CAS, Beijing 100049, China}\\
\\
\footnotesize{$^\ast$Corresponding author; E-mail:  ljchen@nao.cas.cn.}
}

\date{Jul. 1, 2019}

\begin{document}


\baselineskip18pt


\maketitle


\begin{sciabstract}
The Ultra-Long Wavelength (ULW) regime of longer than 10 m (corresponding to frequencies below 30 MHz), remains as the last virtually unexplored window in radio astronomy, and is presently attracting considerable attention as an area of potentially rewarding studies. However, the opaqueness of the Earth's ionosphere makes the ULW celestial radio emission very difficult to detect with ground-based instrumentation. The impact of the ionosphere on ULW radio emission depends on the Solar cycle activities and varies with time. In addition, the ULW spectrum region is densely populated by intensive artificial radio frequency interference (RFI). An obvious solution of these problems is to place an ULW radio telescope in space. However, this solution is expensive and poses non-negligible technological challenges. An alternative approach is triggered by recent studies showing that the period of post 2020 will be most suitable for exploratory ground-based ULW radio observations due to the expected 'calm' state of the ionosphere; the ionospheric cutoff-frequency could be well below 10 MHz, even in the day time. In anticipation of this upcoming opportunity, we propose and present in this paper a concept of an experimental ULW radio array, with the intention of setting it up in Inner Mongolia, China. This ULW facility will use the infrastructure of the currently operational Miangtu Spectra Radio Heliograph (MUSER). The proposed ULW array covers the frequency range from 1 to 72 MHz. This experimental array will be used for exploratory studies of celestial radio emission in the ULW range of the spectrum.
\end{sciabstract}


\section{Introduction}
\label{intro}
The ULW regime, remaining as the last virtually unexplored region of the whole electromagnetic (EM) spectrum, is the most appropriate regime to study several astronomical sciences such as the dark ages, the extragalactic sources, the galactic interstellar media, transients and variable sources, etc~\cite{heino2009},~\cite{boonstra2016}. Due to the strong absorption and scattering of the Earth's ionosphere, the radio emission below 30 MHz are poorly accessible with the ground-based radio telescopes. To date, there are several large Earth-based low frequency radio telescopes that have been built, such as the Low Frequency Array (LOFAR) in Europe operating in the frequency range from 10 to 240 MHz \cite{haarlem2013}, the Long Wavelength Array (LWA) in New Mexico, USA operating in the frequency range from 10 to 88 MHz \cite{ellingson2009}, Giant Ukrainian Radio Telescope in Ukraine operating in the frequency range from 8 to 80 MHz \cite{konovalenko2016}, and the Murchison Wide-Field Array (MWA) in Western Australia operating in the frequency range from 80 to 300 MHz \cite{tingay2013}; However, most of these instruments work above 10 MHz. Besides, the ubiquitous strong artificial radio frequency interference (RFI) also severely limits the observations below 30 MHz in radio astronomy.

In the present situation, a space-based radio telescope operating beyond the Earth's ionosphere appears to be the only viable solution for the ULW astronomy. Some precursor ULW space-borne instruments revealed that the Moon can act as a good shield for the strong RFI from the Earth to provide an ideal radio environment~\cite{alexander1975}; In addition, the intensity of terrestrial radio interferences will also decrease to an acceptable level if the instrument locates far enough (\(>\sim\)157 Earth radii) away from the Earth~\cite{kaiser1996}. Therefore, the feasible way to explore the ULW spectral domain with minimum RFI is that the facility must be outside the Earth's ionosphere, either somehow shielded from the Earth-originated RFI, or located sufficiently far away from the Earth. Up to now, several facilities have been already launched to perform the ULW observations~\cite{herman1973},~\cite{kaiser1996},~\cite{gopalswamy2014}, including the latest radio instrument--the Netherlands-China Low Frequency Explorer (NCLE) \cite{boonstra2017}. However, all of them are mostly single element instruments. There are many concepts of space-borne multi-element interferometers that have been proposed ~\cite{nieuwenhuizen1990},~\cite{jones1998},~\cite{bergman2009},~\cite{bentum2010},
~\cite{saks2010},~\cite{baan2013},~\cite{jones2015},~\cite{boonstra2016},~\cite{rajan2016}, though all of them are still being pursued and in the developmental phase. Considering the complexity in technology and engineering, it is actually quite challenging to achieve these concepts. Thus, given the complexities, engineering challenges, and the very high cost to develop such space-based interferometers, we propose a concept of an experimental ULW radio array at a low cost (around 0.20 Million US dollars (M \$)).

The F-region of the ionosphere is primarily produced by the solar EUV radiation. The \emph{foF2}, an ionospheric parameter that is the critical frequency above which the radio waves from the sky would be reflected back by the Earth's ionosphere, shows a good correlation with solar activity parameters such as sunspot numbers \cite{lukianova2011}. The studies of Janardhan et al. (2015) showed a strong correlation between the variations in \emph{foF2} and the smoothed sunspot number (SSN) during the period 1994-2014, covering solar cycle 23. Further, based on this correlation between \emph{foF2} and the sunspot number, the authors reported an assessment of possible impact of a likely grand minimum on terrestrial ionospheric current systems, based on the one-to-one correlation of sunspot number and night time F-region maximum electron density, and claimed that the period post 2020 will be useful for undertaking systematic ground based low-frequency radio astronomy observations, as the night time ionospheric cutoff-frequency could be well below 10 MHz. Normally, the solar minimum lasts 1-2 years, but the Solar cycle 24 minimum may be extended beyond 2 years. It would therefore provide an excellent opportunity to do ULW astronomy at this special time window, post 2020 for a period of 2-3 years, using our proposed experimental ULW radio array. It will be built in a radio-quiet location at a low cost (around 0.20 M \$) by using available hardware facilities of a solar radio synthesis array (Mingantu spectral radio heliograph, MUSER) \cite{yan2016}. Although there are still strong RFIs existing, by using RFI mitigation methods this experimental radio array will conduct radio observations at low frequencies, and will also be used to study some related technologies for future space-based ULW radio array.

The concept of the radio array is briefly described in Section 2. In Section 3, the system specifications are discussed. In Section 4 the concept design of the whole system  is presented. System analysis is provided in Section 5, followed by discussions on Scientific Objectives achievable from this experimental array. Summary and conclusion are provided in Section 7.

\section{System concept}
\label{sec:concept}
For an ULW radio array, it is impossible to build the antenna element as large as the wavelength scales since its wavelength is long, especially at very low frequencies. To meet the requirements of sensitivity and bandwidth, an active antenna is needed for this kind of array, which has been widely used in modern low frequency radio array systems for ground-based facilities \cite{rhode2001}. The baseline of the array should be long enough to have a reasonable spatial resolution. Besides, the array system should be simple to be within an appropriate budget and to have less construction burdens. Considering these specific requirements, if we set up many low frequency antennas over a big area to form a radio array like LOFAR or LWA, it will be almost impossible to finish the whole system construction (including the antenna implementation, signal transmission, and data receiving and processing) within a very limited budget before 2020. Therefore, we need to have a smart concept to build such kind of a low cost array.

In Inner Mongolia in China, a solar-dedicated radio interferometer, MiangtU SpEctra Radio Heliograph (MUSER), is in operation. It can carry out imaging spectroscopy of the Sun, to produce high spatial, high time and high frequency resolution images of the Sun at different frequencies simultaneously. MUSER has two arrays with parabolic dishes operating at different frequency bands, such as MUSER-I, the low frequency array, operating at frequencies from 0.4 to 2 GHz and MUSER-II, the high frequency array, operating at frequencies from 2 to 15 GHz. The MUSER-I contains 40 antennas of 4.5 m diameter each, and the MUSER-II contains 60 antennas of 2 m diameter each. All the 100 antennas of MUSER array are located on three log-spiral arms with the longest baseline being about 3 km. The radio frequency (RF) signal detected by each MUSER antenna is transmitted through optical fibers to an indoor facility containing analog and digital receivers.

\begin{figure}[t]
\centering
\includegraphics[width=\columnwidth]{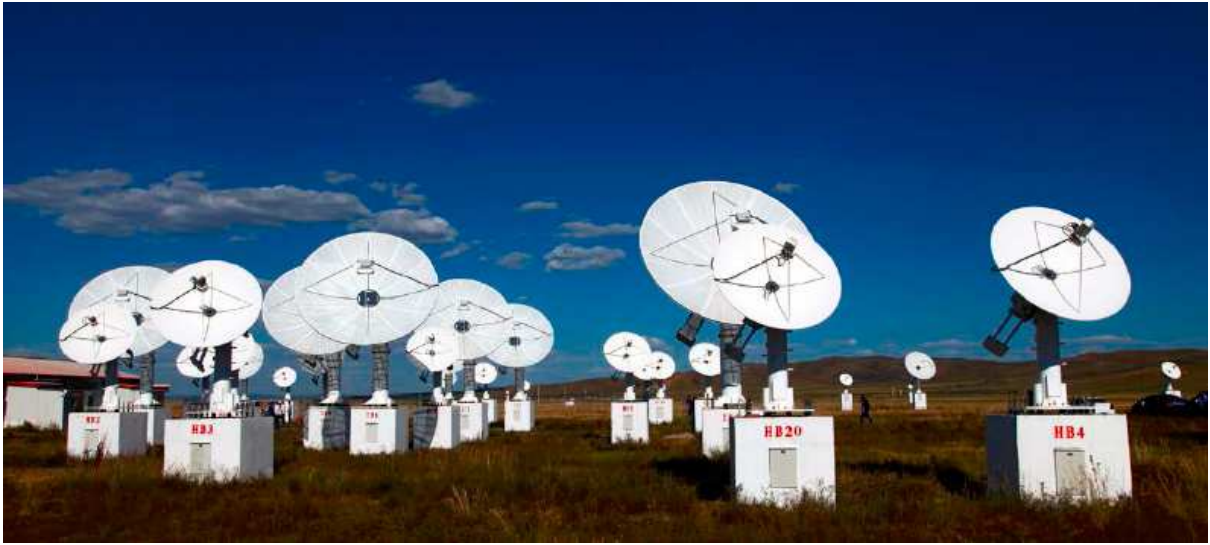}%
\caption{The central part of MUSER with 4.5 m diameter parabolic antennas for MUSER-I and 2.0 m parabolic antennas for MUSER-II.
\label{fig:muser}}
\end{figure}

In the design of MUSER, the optical fibers for each antenna are redundant and only half of them are used. The other half are still available. We consider using the available optical fibers to transport the RF signals, which is also called as RF over fiber. The RF signal is close to the optical transmitter and fiber connector in MUSER antenna base so as to avoid any long cable signal transmission. As shown in Figure~\ref{fig:muser}, the antennas of MUSER array are forward-feed dish antenna with their feeds supported by four struts. We design the antenna of the proposed experimental radio array as a thin copper wire and fix it along the feed support strut, with which we can get a cross-dipole antenna by combining the four wires on the support struts. The low noise amplifier (LNA) of this antenna is fixed on the backside of the front-end box of MUSER antenna. Considering the antenna of MUSER-II array is only 2.0 m in diameter, it is not big enough to support a reasonably big cross-dipole antenna. So we prefer to set up the cross-dipole antenna only on MUSER-I dishes, which allows us to have maximum 40 antennas for the proposed ULW radio array. The RF signals detected by the antennas will be transported by the fibers to an indoor digital receiver, which is employed to convert the radio frequency signals to digital signals, and to perform the signal processing. Besides, a hard disk array is used to store all of the observational data for further scientific studies.

As mentioned above, since we set up the dipole antennas on MUSER dish antennas for the proposed experimental ULW radio array, we do not require extra support mechanisms and locations to set up the antennas. Furthermore, the dipole antenna can share the same pointing control system with MUSER together, which will make the dipole even have a flexible pointing. With the available optical fibers, we can easily solve the problem of long-distance signal transmission between the antenna and the indoor receiver. The only thing required is to develop an optical transceiver module. The digital receiver is a key instrument and is needed to be carefully designed for this experimental ULW radio array.

\section{System specification}
\label{sec:spec}
Based on the concept introduced in Section \ref{sec:concept}, we propose an experimental ULW radio array with the main specifications as shown in Table \ref{tab:spec}.
\begin{table}
\caption{Basic specifications of the ULW radio array}
\label{tab:spec}       
\begin{tabular}{ll}
\hline\noalign{\smallskip}
Antenna  & 5 m, \(36 \times 2 \) dipoles \\
Polarization & Dual linear \\
Frequency range & 1\(\sim\) 72 MHz \\
Frequency channel & 1 \(\sim\) 16  \\
Frequency resolution & \(\sim\) 73 kHz  \\
Spatial resolution & \(0.1^{\circ}\)@ 72 MHz \(\sim 7.0^{\circ}@1 MHz\) \\
Sky noise limited & 10 dB, \(7.5 \sim\) 51 MHz \\
ADC & 12 bits @ 160 MHz \\
Simultaneous beam & \(\geq\) 8 \\
System dynamic range & \(\sim\) 36 dB  \\
\noalign{\smallskip}\hline
\end{tabular}
\end{table}

In order to make radio observations beyond the ULW regime, the upper limit of operating frequencies of this ULW radio array is set to higher than 30 MHz so that joint observations can be performed with other low frequency radio telescopes for calibration purposes and science studies. Some antennas of MUSER-I array are located close to two small villages, so only 36 antennas of MUSER-I array are used to set up the cross-dipole antennas for the ULW array so as to avoid the possible RFIs. As the longest baseline of the ULW array is around 3 km, the spatial resolution varies from \(0.1^{\circ}\) at 72 MHz to \(\sim 7.0^{\circ}\) at 1 MHz. Simultaneously, more than eight beams will be formed to do the radio observations. For each single antenna system, the system dynamic range is about 36 dB relative to the  galactic background.

\begin{figure}[h]
\centering
\includegraphics[width=\columnwidth]{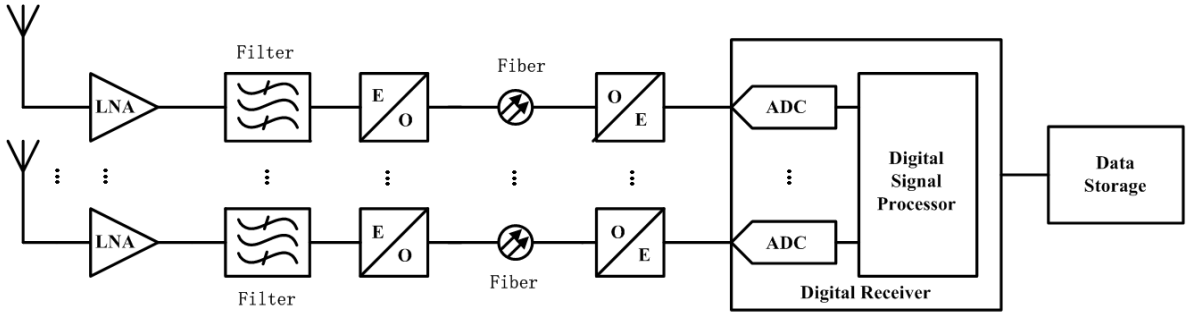}%
\caption{System schematic of the experimental ULW radio array.
\label{fig:system}}
\end{figure}

\section{System description}
\label{sec:design}
The whole system of this experimental radio array mainly includes the active antennas, the optical modules and fibers, and a digital receiver. A systematical schematic is shown in Figure \ref{fig:system}. Using the RF over fiber, the radio signals sensed by the active antennas will be transported to the indoor digital receiver with fibers. The digital receiver will then sample the electrical signals and process them with the designed algorithms for scientific outputs.

\subsection{Antenna}
\label{subsec:antenna}

\begin{figure}[h]
\centering
\includegraphics[width=3.2in]{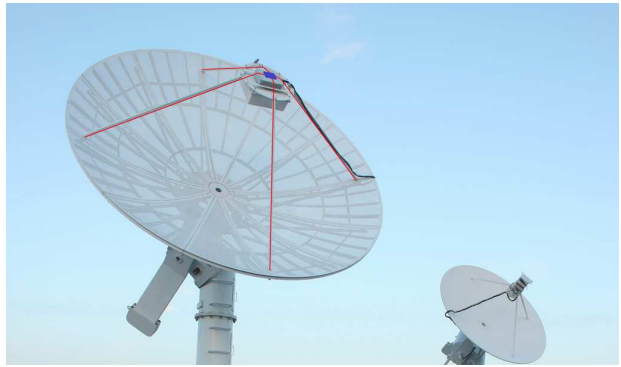}%
\caption{Antenna setup of the experimental ULW radio array. The red lines on the feed support struts are two dipole antennas with inverted V shapes, the LNA box (blue part) is fixed on the back side of the front box of the dish antenna.
\label{fig:antenna}}
\end{figure}

For a radio array, the antenna is a key element of the whole system. At low frequencies, an active antenna is the inevitable choice to obtain a good gain and a wide frequency band. As described in Section \ref{sec:concept}, the antenna of the proposed experimental ULW array will be set up on MUSER-I's parabolic antenna as shown in Figure \ref{fig:antenna}. By placing four copper wires on the feed support struts, we form two dipole antennas with an inverted V shape to detect the radio signals. Each support strut of the parabolic antenna is \(\sim \)2.5 m long, so the maximum length of each dipole antenna is \(\sim \)5.0 meters. The antenna is isolated from the strut by its cable insulation. The angle between one pair of monopoles is about \(106^{\circ}\), which in principle will broaden the radiation pattern of the dipole antenna to observe almost all of the sky. The parabolic antenna can be considered as a ground plane to the dipole antenna, and influence the antenna performance, especially at higher frequencies in the operating frequency band.

\begin{figure}[h]
\centering
\includegraphics[width=\columnwidth]{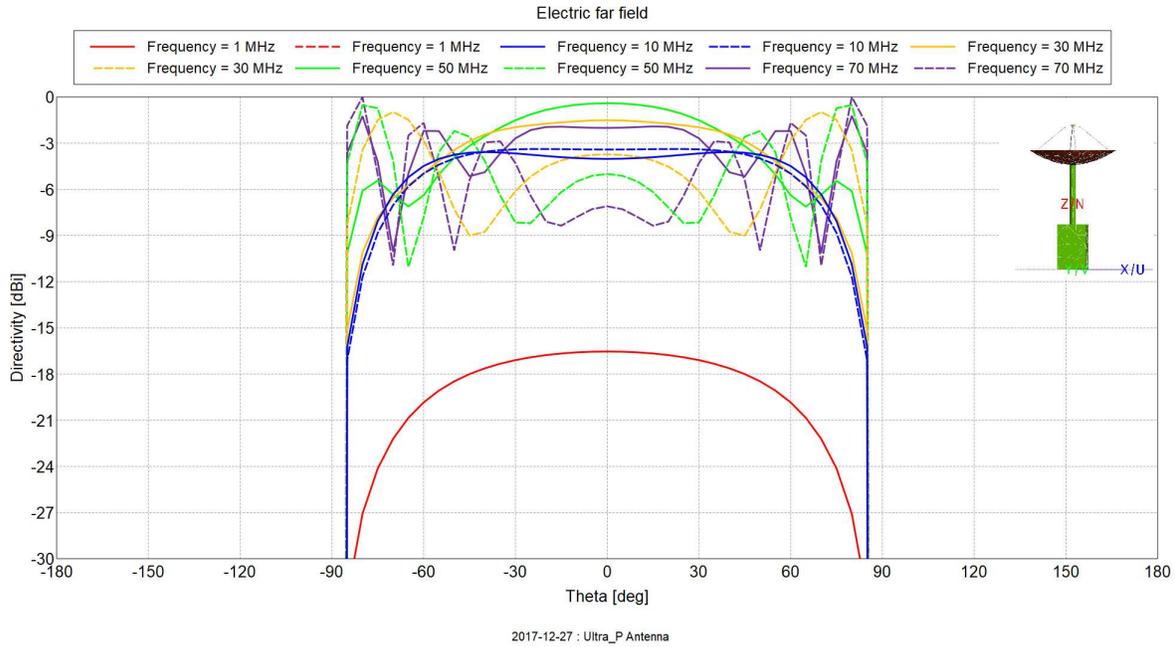}%
\caption{Radiation pattern simulations of dipole antenna by FEKO at different frequencies. The solid lines and dashed lines represent the dipole antenna patterns with and without the parabolic antenna, respectively. The different colors denote different frequencies from 1 MHz to 70 MHz.
\label{fig:simulations}}
\end{figure}

As shown in Figure \ref{fig:simulations}, the parabolic antenna improves the dipole antenna radiation pattern by suppressing the sidelobes at most of the frequencies. Detailed simulations show that the antenna patterns will become saddle-shape patterns between 10 and 20 MHz. A further analysis reveals that it can be ascribed to the reflection influence by the ground. Although the parabolic antenna can shield the radiations from the ground more or less, the radiation pattern of the dipole antenna will still be affected by the ground reflection since the dish size is quite small relative to the wavelength. While tracking the radio sources the dipole antenna moves with the parabolic antenna in order to point in different directions, so its radiation pattern will change inevitably with the orientation of the parabolic antenna.

\begin{figure}[h]
\centering
\includegraphics[width=\columnwidth]{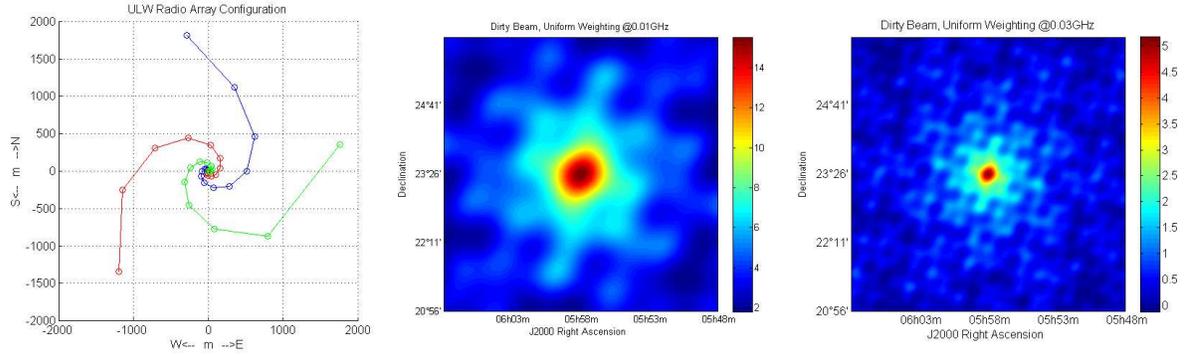}%
\caption{Array configuration and synthesis beams in zenith direction. Left: Proposed ULW array configuration. Middle: Array synthesis dirty beam at 10 MHz. Right: Array synthesis dirty beam at 30 MHz.
\label{fig:array}}
\end{figure}

In the implementation, we will optimize the antenna length to obtain a good balance among the different antenna performances including the antenna pattern, frequency band, gain and efficiency.

All the antennas of the proposed ULW array are located on three log-spiral arms along with the MUSER's antennas together, as shown in Figure \ref{fig:array}. The maximum baseline of this ULW array is 3287 meters, which will provide a spatial resolution of around \textbf{\emph{\(7.0^{\circ} \sim 0.1^{\circ}\)}} at the frequency range from \(1 \sim 72\) MHz. The array synthesis beams are simulated as plotted in Figure \ref{fig:array}.

\begin{figure}[h]
\centering
\includegraphics[width=\columnwidth]{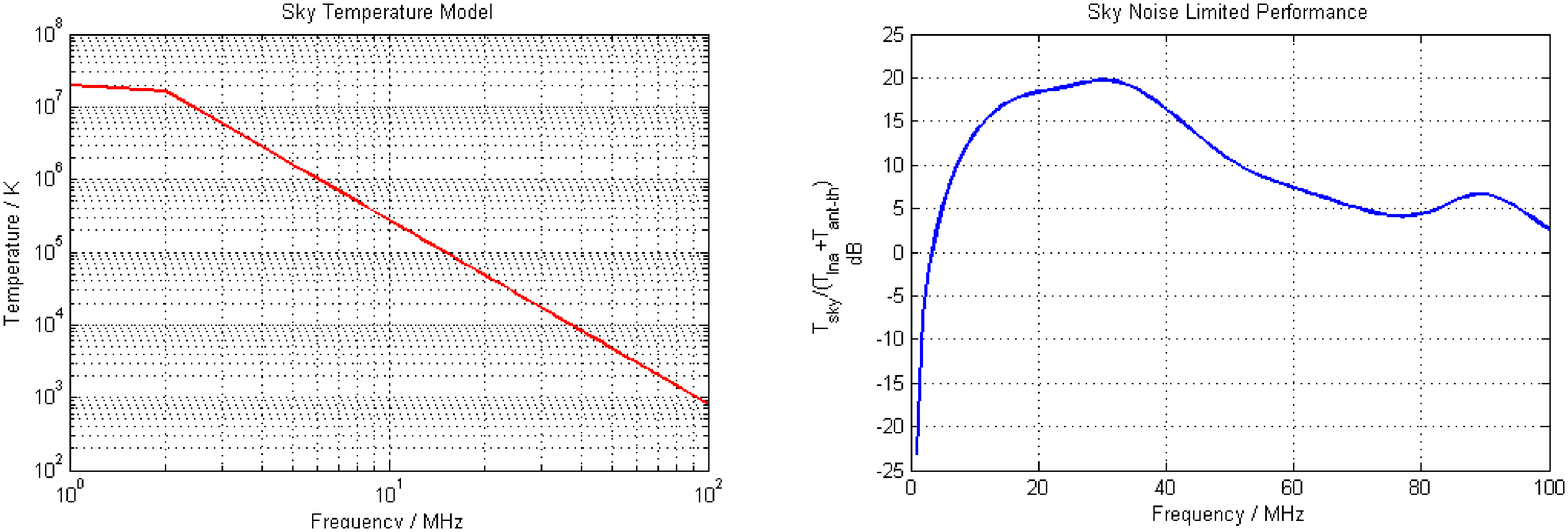}%
\caption{Simulations of the sky noise limited performance. It is evaluated by the ratio between the detected sky noise~\(T_{sky}\) and the receiver noise~\(T_{lna}+T_{ant_thermal}\) of the antenna system. A high ratio is essential for the array to carry out the synthesis imaging with a high spectral resolution. Left: The sky temperature model~\cite{heino2009} used in the simulations. Right: The simulations of the sky noise limited performance for the designed active antenna.
\label{fig:snr}}
\end{figure}

Behind each dipole antenna, a low noise amplifier is used to amplify the detected signals with low noise and to obtain a matching between the antenna and transmission cable. This LNA includes two stage amplifiers. The first stage is a voltage amplifier with high input impedance, and the second stage is a current feedback amplifier~\cite{chen2018}. The available gain of the active antenna is low due to the mismatching between the dipole antenna and LNA. In 50-ohm system the noise figure of the LNA is around 2.0 dB over most of the operating frequency band. This LNA is designed for the dipole antenna above to achieve the sky noise limited performance within the operational frequency band~\cite{chen2018}. With a sky temperature model~\cite{heino2009}, we simulate the sky noise limited performance of this active antenna. As shown in Figure \ref{fig:snr}, it can achieve a 10 dB sky-noise limitation between 7.5 and 51 MHz, and a 5 dB sky-noise limitation between 5 and 70 MHz. Two LNAs are combined together on one printed circuit board (PCB), which is assembled in a LNA box. This box is fixed on the front-end receiver box of MUSER-I as shown in Figure \ref{fig:antenna}.

\subsection{Optical transceiver \& fiber}
\label{subsec:optical}

\begin{figure}[h]
\centering
\includegraphics[width=\columnwidth]{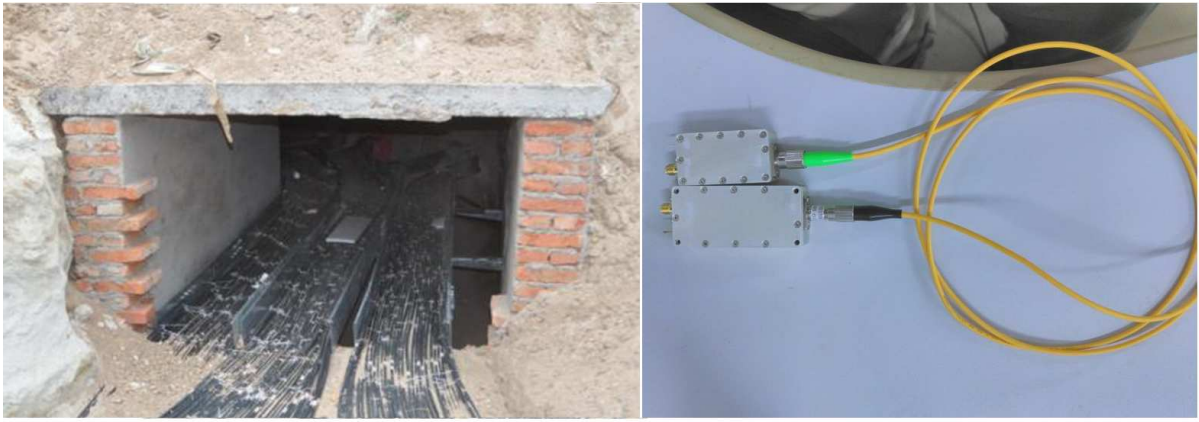}%
\caption{Optical fiber and transceiver. Top-left: Optical fiber underground. Top-right: Optical transceiver prototype. Bottom: Noise figure measurements of the transceiver.
\label{fig:fiber}}
\end{figure}

As mentioned above, we use the design of RF over fiber to transmit radio signals to indoor receivers. The radio signal detected by the active antenna will be transmitted to an optical transmitter via a 10-meter coaxial cable, which will introduce an extra noise temperature of \(\sim 7-28\) K. The optical transmitter in the MUSER's antenna base converts the radio signal to the optical signal. Then, the redundant fiber of MUSER is used to transmit the optical signal to the indoor receiver, where it will be converted back to the radio signal by an optical receiver. To decrease the temperature influence on phase differences between the different antennas, all of the optical fibers of MUSER are buried 2.5-meter deep underground covering the maximum baseline length of \(\sim 3\) km. The optical transceiver including a transmitter and a receiver has a total gain of \(\sim 30\) dB, which allows the radio signal to be sampled by the Analog to Digital Converter (ADC) at a proper power level. As the available gain of the active antenna is low, we need to decrease the system noise temperature in order to improve the sensitivity of the receiving chain. Thus, this transceiver is designed to have a low noise figure of about \(\leq 2.0\)dB in most of the frequency band, as shown in Figure \ref{fig:fiber}.

\subsection{Digital receiver}
\label{subsec:digital}
For a low frequency radio array, the radio signal detected by the antenna will be sampled directly by a digital receiver without passing through a complex analog receiver. In this way, almost all the signal processing is done in the digital domain, thus the digital receiver is the core instrument of the whole array system. In the proposed experimental ULW radio array, the digital receiver will sample the antenna signal with a 12-bit ADC operating at 160 Million Samples per Second (MSPS). The time delays between different antennas will be compensated first to the sampled different antenna signals. Then a polyphase filter bank (PFB) combined with a 2048-point fast fourier transform (FFT) will be used to channelize these signals, in which 1024 sub-band signals will be obtained with a bandwidth of \(\sim78\) kHz. Each sub-band signal will be compensated for the phase differences between the different signal receiving chain. During observations, we may select at most 16 frequencies to do further scientific processing with a flexible bandwidth by integrating the adjacent sub-band signals. These 16 frequencies can be almost random across the whole frequency band for each observation, by changing them all the frequencies can be covered with a longer time.

\begin{figure}[h]
\centering
\includegraphics[width=\columnwidth]{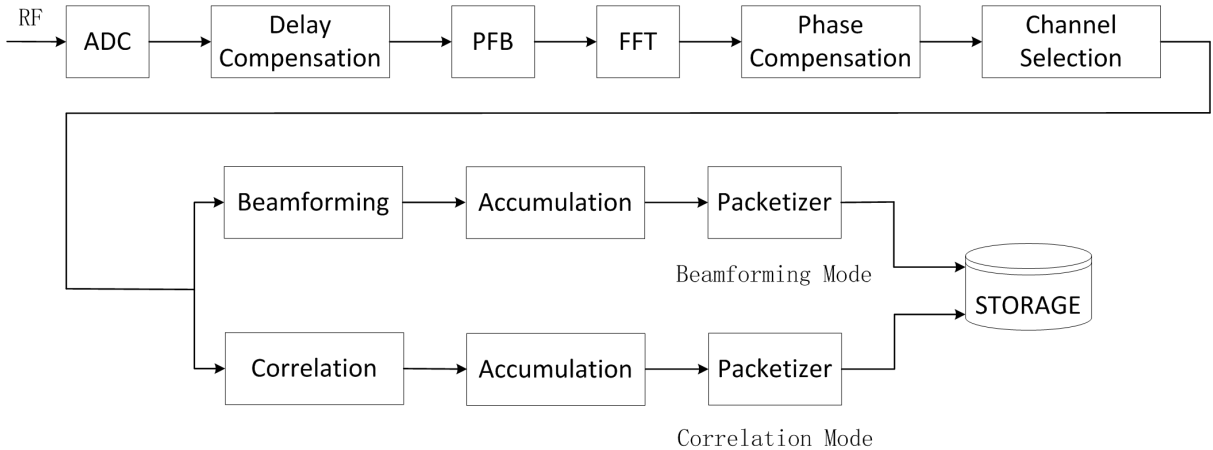}%
\caption{Signal flow in the digital receiver. The signal processing is carried out in two different methods for different observation modes. Note that this block diagram just shows the signal processing flow, and is not the exact design scheme of the digital receiver.
\label{fig:digtal}}
\end{figure}

In order to observe different radio sources in the sky, the proposed experimental ULW radio array is designed to work in two different scientific modes, multi-beamforming mode and synthesis imaging mode. Therefore, the signals are processed further in two different ways in the digital receiver as shown in Figure \ref{fig:digtal}. In the beamforming mode, all the antenna signals will be multiplied by the complex weighting coefficients and then summed together to get an output with a specified beam. Multi-beam signals can be obtained simultaneously by using different groups of weighting coefficients in parallel.  In the design, the beamforming for broad-band radio signals can be achieved by integrating the adjacent narrow-band beamforming signals. In the synthesis imaging mode, all the antenna signals with two polarizations will be cross-correlated to measure four Stokes parameters that will be used for imaging celestial radio sources. The beamforming mode and the synthesis imaging mode can work together or separately. Both of them have the integration time of 10 ms, the maximum data rate is around 32 MB/s.

\section{System analysis}
\label{sec:analysis}
In this experimental ULW radio array, the induced radio signals on the antenna by celestial radio emissions will be transmitted to the indoor receiver for digitization and further signal processing. To analyze the noise distribution, a reference plane is set between the antenna and the LNA. Before this reference plane, the noise is called the antenna noise, which includes the noise from the sky and the thermal noise of the antenna itself. According to the temperature model in~\cite{heino2009}, the different noise contributions in the system have been simulated. The average sky noise temperature is around \(7.1\times10^{5}\) K within the operating frequency band. The antenna only detects part of the sky noise, which is determined by the antenna efficiency~\cite{chen2018}. The average antenna noise temperature is around \(8.7\times10^{4}\) K. After the reference plane, the noise is called receiver noise, which includes the LNA noise, the optical transceiver noise and the digital receiver noise. The average LNA noise at the output is around \(2.1\times10^{4}\) K, the noise temperature of the optical transceiver is around 170K at the input. As the gain of the transceiver is as high as 30 dB, the equivalent noise of the digital receiver at the transceiver input can be ignored. Thus, the total receiver noise at the LNA output can be calculated as the sum of the LNA noise and the optical transceiver noise, which is about \(2.1\times10^{4}\) K. The available gain of the active antenna is about 7.8 dB, so the equivalent receiver noise at the reference plane can be obtained by dividing by the available gain, it is ~\(3.6\times10^{3}\) K. The total system noise is equal to the sum of the antenna noise and the receiver noise, which is about~\(9.1\times10^{4}\) K.

\begin{figure}[h]
\centering
\includegraphics[width=\columnwidth]{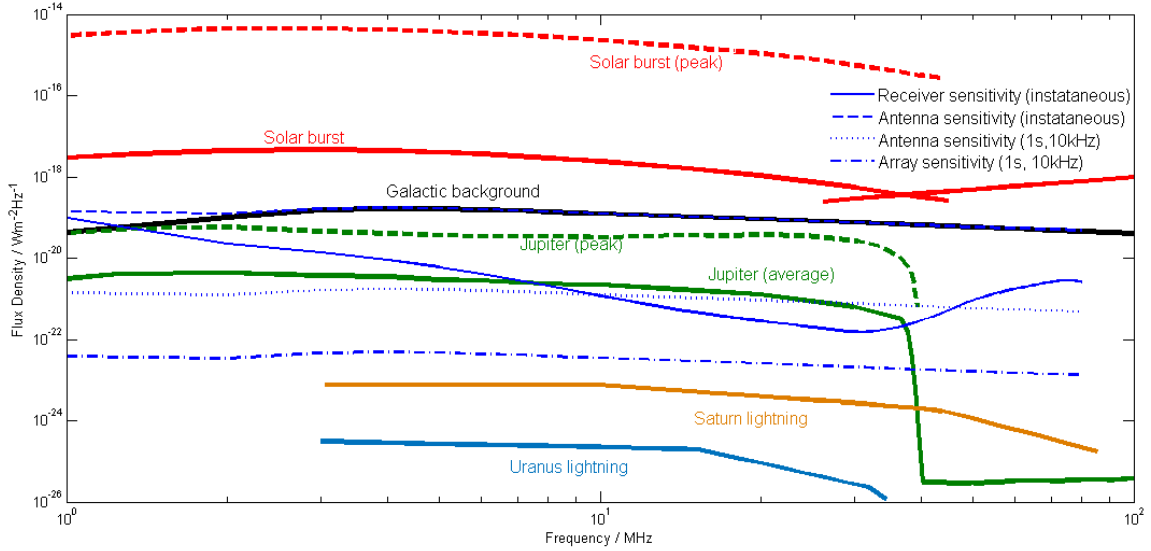}%
\caption{Antenna and array sensitivities comparing with the flux density of different radio emissions. The different radio emissions are replotted with the data in \cite{zarka2012}. Note that the receiver sensitivity is calculated as the antenna system sensitivity without taking into account the galactic background noise. The sensitivities of the antenna and array are simulated with different integration times and bandwidths.
\label{fig:flux}}
\end{figure}

Based on the noise contribution analysis, we simulated the sensitivities of the antenna and array system. As shown in Figure \ref{fig:flux}, if the galactic background noise is excluded from the system noise, the antenna sensitivity (named as receiver sensitivity here) is much lower than the galactic background over most part of the operating frequency band. This is also shown in Figure \ref{fig:snr}. With a certain integration time and observing bandwidth, even one single antenna will be sensitive enough to detect both the peak and the average Jupiter radio emissions. At the ULW band, the solar bursts are strong enough, so both the antenna and array will be able to detect them.

For a low frequency radio array, if the antenna signal is sampled directly by the digital receiver, the dynamic range of the system will only be determined by the LNA and the digital receiver. Normally, a good LNA will have a sufficient dynamic range to amplify the radio frequency signal since it is very small in radio astronomy. In that case, the system dynamic range will mainly depend on the digital receiver. For the proposed experimental ULW radio array, the optical transceiver is used to transmit the radio signal between the antenna and the digital receiver, which will certainly affect the system dynamic range. Suppose there is no RFI, the system simulations show that the total power of the detected radio signal on the antenna is about -56.6 dBm. The gain of the optical transceiver is around 30 dB and the 3 km optical fiber attenuates the signal by ~\(1\) dB, then the signal power at the ADC input will be about -27.6 dBm. For a 12-bit ADC with a full-scale of 10 dBm, this power will fill 5 bits, the other bits will be left for the high-dynamic range signals. However, the input power of the 1 dB compression point for the optical transceiver is around -22 dBm, so its dynamic range will be limited to \(-22-(-56.6)=36.6\) dB. If we take into account both of the optical transceiver and the digital receiver, it can be concluded that the system dynamic range will be around 36 dB without considering the RFIs.

\begin{figure}[h]
\centering
\includegraphics[width=\columnwidth]{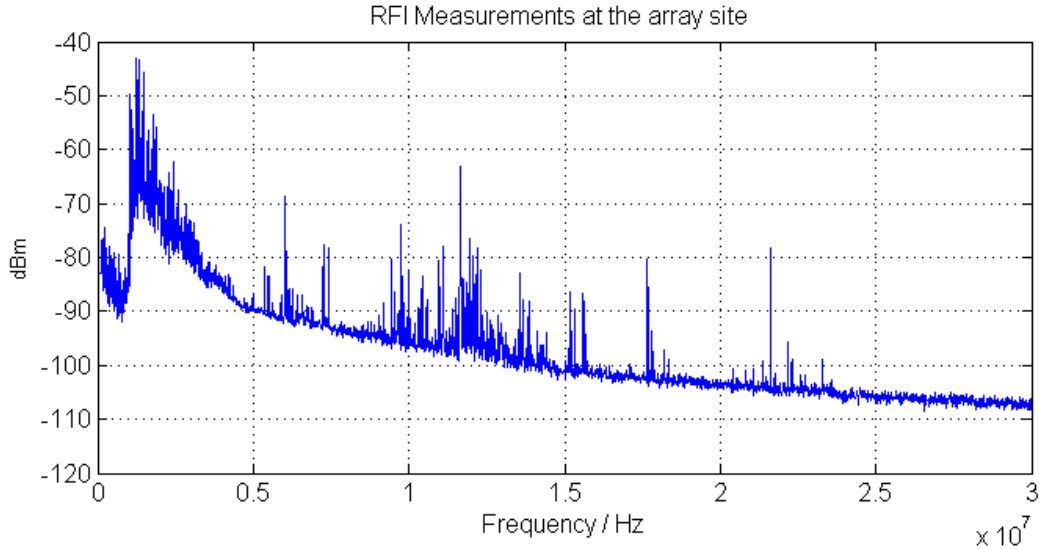}%
\caption{RFI measurement below 30 MHz with a 2.5-meter antenna at the experiment radio array site on Jan. 3, 2019. The antenna is set up with the similar configuration as shown in Figure \ref{fig:antenna}, not on the dish, but 1.5 meters high above the ground. The measurement has a spectral resolution of 10 kHz.
\label{fig:rfi}}
\end{figure}

Although the proposed experimental radio array is located at a radio-quiet zone, there are still many RFIs at the ULW regime, as shown in Figure~\ref{fig:rfi}. It is a serious problem for the radio observations within this frequency range. In this case, different techniques will be studied (including flagging, filtering, thresholding, spatial nulling, machine learning, etc) to mitigate the RFIs. Some measures have already been taken at the array site, such as the RFI monitoring, the study of the RFI mitigation algorithms in the data processing, etc. Besides, the ionosphere calibration is another issue for the low frequency radio array. Compared with the RFI problem, it is a little easier to solve this issue since all the antennas of the proposed experimental array are located within only 3 km, and the impact of the ionosphere in this distance range is less~\cite{mevius2016}. Even then, we will still study how the ionosphere affects the radio observations with this ULW array.

\section{Expected scientific studies}
\label{sec:science}
Based on the concept design mentioned above, the proposed experimental ULW radio array will be a competent instrument capable of studying some astronomical objects and phenomena similar to the science cases as described in ~\cite{heino2009}. A few of them include studying solar radio bursts, mapping the radio sky at ULW regime, investigating planetary radio emissions, and trying the VLBI observations with other low frequency facilities.

Despite the great importance to space weather, the intense radio bursts associated with solar flares and coronal mass ejections (CME) are poorly understood on their physical mechanisms, with the consequence that neither accurate models nor reliable prediction tools exist. Among these solar radio bursts, two main types are Type II bursts which drive energetic shocks through the solar corona and interplanetary medium (~1000-2000 km/s), and Type III bursts which result from mildly relativistic (~0.1-0.3 c) electron beams propagating through the corona and interplanetary space. At the ULW regime, the proposed experimental array will be able to detect and track the radio bursts in the interplanetary space within about 10-20 solar radii. Combined with other solar radio instruments, such as MUSER, and the Interplanetary Scintillation (IPS) telescope and meter-wave spectral radio heliograph that are planned to be built at Mingantu observation station, the proposed ULW array will provide a unique opportunity to study the propagation and evolution of the interplanetary CME and non-thermal high-energy particle beams.

In radio astronomy the ULW range remains as the last virtually unexplored region of electromagnetic spectrum. Due to the limitation of the terrestrial radio facilities, so far the best available maps of the low-frequency sky have a resolution of 5 degrees at 10 MHz with a very poor dynamic range \cite{cane2001}. At the frequencies below 10MHz, the only view of the sky we have until now is made by the Radio Astronomy Explorer 2 (RAE-2) satellite in 1973 \cite{novaco1978}. However, it provides hardly any information about the individual astronomical source. During the period of post 2020, the proposed experimental radio array will provide a high possibility of mapping the sky at ULW regime with a reasonable spatial resolution.

Among planetary radio emissions, the Jupiter emissions are intense (MJy), point-like, \(100\%\) polarized, quasi-permanent in the hectometer-kilometer (0.3-3 MHz) range, and reasonably predictable in the decameter (30 MHz) range \cite{zarka2004},~\cite{nigl2007}. Jupiter can be considered as an ideal calibration source and a perfect target for the ground-based long wavelength array. For Saturn, the proposed experimental ULW array may permit seasonal or secular effect studies of its radio emissions, as well as for Uranus. Quasi-continuous observations for all planets will reveal the time-variability of the emissions related to planetary rotation periods, variability of the solar wind, and satellite modulations.

The Netherlands-China low frequency explorer (NCLE) is a radio instrument onboard the Chinese Chang'E 4 lunar mission \cite{boonstra2017}. In the Earth-Moon L2 orbit it can perform the astronomical observations at the frequency range of 0.08 - 80 MHz. We are deeply involved in this project. The proposed experimental ULW array will make some joint radio observations with NCLE together.

\section{Summary, conclusions and future work}
In this paper, an experimental ULW radio array has been proposed with a relatively easy implementation and at a relatively low cost. The preliminary design of the whole system has been introduced as well. Simulations show that the antenna system of the ULW array is sky noise limited over most part of the operating frequency band. By using the RF over fiber design, the antenna signal is transmitted over a long distance with low attenuation and phase variation. Also, we show that the array can achieve reasonable sensitivity and dynamic range. Several scientific studies can be pursued using the proposed array, as well as the array can be used as a precursor for the future space-based ULW radio array.

Currently, all of the sub-system developments of this experimental array have almost been finished. Next they will be assembled and tested soon. The preliminary observations are expected to be done by the end of 2019. In addition, we will also optimize the signal processing algorithms of the digital receiver further to obtain more flexibilities and better performances. The RFI mitigation techniques will also be studied, as well as the ionosphere calibration methods.

\section*{Acknowledgements}
This work was supported by the grant of National Natural Science Foundation of China (NSFC) 11573043, 11203042,11790305,11433006, and the regulation grant of National Astronomical Observatories, Chinese Academy of Sciences. We had the big benefit of valuable discussions with Prof.S.Ananthakrishnan from Pune, India during his visit to Miangtu in 2014 when this idea was brought up. The authors thank Dr. Susanta Kumar Bisoi from NAOC, CAS, Beijing for the discussion and help. We also thank the anonymous referees for the valuable comments and suggestions.

\bibliographystyle{spbasic}      


\end{document}